\documentclass[11pt,a4paper]{article}

\usepackage{amsmath,epsfig,amssymb,amsthm,a4wide,longtable,multicol,lineno,color,relsize,fullpage}

\everymath{\displaystyle}

\newtheorem{thm}{Theorem}
\newtheorem{rem}[thm]{Remark}

\def\it{\textit} 
\def\bf{\textbf} 
\def\tt{\texttt} 
\def\ul{\underline}

\def\<_m{<_{\mathrm{m}}}

\newcommand{\be}{\begin{equation}}
\newcommand{\ee}{\end{equation}}
\newcommand{\benum}{\begin{enumerate}}
\newcommand{\eenum}{\end{enumerate}}
\newcommand{\bit}{\begin{itemize}}
\newcommand{\eit}{\end{itemize}}

\title{Maximum independent set (stable set) problem: 
Computational testing with binary search
and convex programming using a bin packing approach}

\author{Prabhu Manyem (Retired)\thanks{Currently living in Adelaide, Australia, after retiring.}\\
College of Science\\
Nanchang Institute of Technology\\
Nanchang 330099 China\\
\tt{prabhu.manyem@gmail.com}}

\begin{document}

\maketitle{\thispagestyle{empty}}

\begin{abstract}
This paper deals with the maximum independent set (M.I.S.) problem, also
known as the stable set problem.  The {basic} 
mathematical programming model that captures this problem is an Integer
Program (I.P.) with zero-one variables $x_j$ and only the \it{edge
inequalities} with an objective function value of the form
~$\textstyle \sum_{j=1}^N x_j$~ where $N$ is the number of vertices in
the input.
We consider $LP(k)$, which is the Linear programming (LP) relaxation of
the I.P. with an additional constraint $\textstyle \sum_{j=1}^N x_j  = k$
(0 $\le k \le N$).
We then consider a convex programming variant $CP(k)$ of $LP(k)$, which
is the same as $LP(k)$, except that the objective function is a nonlinear
 convex function (which we minimise).
 ~The M.I.S. problem can be solved by solving $CP(k)$ for
every value of $k$ in the interval ~0 $\le k \le N$ where the 
convex function is 
minimised using a \it{bin packing} type of approach.
In this paper, we present efforts to developing a convex function for
$CP(k)$; in particular, a convex function that is polynomial (in Section 
\ref{sec:polyConvexFunc}).  When we provide partial solutions to an
instance for 5 vertices (out of 150), the frequency of hitting an optimal
complete integer solution increases significantly.

\end{abstract}



\bigskip

\section{Recent updates}

\it{All files moved to the google drive folder named ``Convex optimisation
method"}:

\texttt{https://drive.google.com/drive/folders/1t6aUcniJE0xwGwlBhrzxgzZzZM9QHHlq}

The latest updates (as of March 2023) are in Section
\ref{sec:polyConvexFunc}, developing and testing a polynomial convex
function.

March 31 2023:  Added Section 3.6.3 and Appendix C.

April 6:  Minor changes to Sec. 3.6.4 where a 150-vertex instance is
solved.

May 2:  Added Sec. 3.6.5 with results from 2 instances with 53 vertices each.

May 4:  Minor changes to Sec. 3.6.5 and Sec. 3.6.6.

May 22: Added Sec. \ref{sec:oeis_1dc_256}, \it{An instance with 256
vertices}.

June 12: Added Sec. \ref{sec:approx}, 
\it{Approximation results from the Literature}.

July 12:  Added Remark \ref{rem:twoStepProc}, expanded Remark 
\ref{rem:twoStepPoints} and the Conclusion at the end of the paper.

August 17: Added Section \ref{sec:recog_integer}, \it{Recognising
an Independent Set from a Fractional Solution}.

Early October, 2023: Added Sections \ref{sec:multipleAMPL} and
\ref{sec:edgeIneq}, and made minor changes to Sec.
\ref{sec:recog_integer}. 

October 22, 2023: Added Paragraph \ref{para:k29Expts} (Page
\pageref{para:k29Expts}) about \newline
 \bf{Experimental results for $k =$ 29, the
optimal solution value} for the 150-vertex instance.

November 16, 2023:  Minor changes to the Conclusion.

December 19, 2023:
Added Sections \ref{sec:N150k29PartialSoln} and 
\ref{sec:150VertexNewFunction}.

\subsection{Definitions}

The decision version of the maximum independent set (M.I.S.) problem,
also known as the stable set problem, is as follows:

Given a constant $K$ and a graph $G = (V, E)$, where $V$ is the set of
vertices and $E$ is the set of edges in $G$, is there a subset $S$ of $V$
such
that (i) no two members of $S$ are adjacent to each other in $E$, and
(ii) $|S|$ (the cardinality of $S$) is at least $K$?

The number of vertices $|V|$ in a graph is denoted by $N$.

The decision version of M.I.S. is known to be NP-complete \cite{gj}.

M.I.S. (that is, its optimisation version), is captured by the following
\it{basic} Integer Program (I.P.).

\bf{Problem 1}.
\begin{equation}
\begin{array}{rcl}
\mbox{Maximise} ~ Z_1 & = & \sum_{j \in V} F[j] \\ 
\mbox{Subject to} & & \\  [2mm]
F[i] + F[j] & \le & 1 ~~ \forall ~ (i,j) \in E \\ [2mm]
F[j] & \in & \{0, 1\} ~~ \forall ~ j \in V.
\end{array}
\label{original_IP}
\end{equation}

Note that for every vertex $j$, $F[j]$ is a binary variable which can be
assigned exactly one of two values (either zero or one).. $F[j]$ = 1 if
vertex $j$ is a member of the Independent set, and zero otherwise.

The constraints of the form ~$F[i] + F[j] \le 1$~ are known as \bf{edge
inequalities}.  

\bf{Number of edge inequalities}:
Since there is one such constraint per edge, the number
of these constraints is polynomial in $N$.

Let \it{OPT (I.P.)} be the optimal solution value of the Integer Program
in Problem (\ref{original_IP}) above, and let OPT (LP) be the optimal solution value for the Linear
Relaxation below in Problem (\ref{original_relaxation}).

The \it{Linear Relaxation} of the above I.P. in (\ref{original_IP}) is
the following Linear Program (L.P.).

\bf{Problem 2}.
\begin{equation}
\begin{array}{rcl}
\mbox{Maximise} ~  Z_2 & = & \sum_{j \in V} F[j] \\ 
\mbox{Subject to} & & \\  [2mm]
F[i] + F[j] & \le & 1 ~~ \forall ~ (i,j) \in E \\ [2mm]
 0 \le F[j] \le 1 &  &  \forall ~ j \in V.
\end{array}
\label{original_relaxation}
\end{equation}

It is known that Linear Programs are in the computational complexity
class \bf{P}; that is, they can be solved in polynomial time; for
example, using algorithms such as the Ellipsoid method or the Interior
Point method \cite{fang1993Linear}. Hence \it{every} instance of Problem
2 can be solved in polynomial time.

Whether we solve Problem 1 or Problem 2,
the underlying input instance is the same,
which is a graph, say $G_1$ = ($V_1$, $E_1$). ~For example, the vertex
set can be $V_1$ = \{a, b, c, d\}, and the edge set can be $E_1$ = 
\{(a, c), (b, d), (a, d)\}.

Suppose we solve an instance $G_k$ of Linear Program \ref{original_relaxation}
in polynomial time (for example, using an Interior Point algorithm); and
suppose in the optimal solution, for every vertex $j$, $F[j]$ is integer
(that is, either zero or one).  This means, we have solved the same
instance $G_k$ for the Integer Program (Problem 1) in polynomial
time as well.  And if we can do this for \bf{every} instance, then we can
conclude that Problem 1 can be solved in polynomial time.

\bf{Maximal Clique}: This is a clique $Q$ that cannot be enlarged by
adding vertices to $Q$.

\bf{0-1 (zero-one) Integer Program}: In these Integer Programs, the
feasible set for every variable $x_i$ is $\{0, ~1\}$. These are also
called \it{binary} integer programs. (Hence in the Linear Programming
relaxation, 0 $\le x_i \le$ 1.)

\subsection{Approximation results from the Literature}\label{sec:approx}

Let $A(G)$ = value of the solution returned by approximation algorithm
$A$ for M.I.S. with graph $G$ as the input instance.

Let $OPT(G)$ = optimal solution value for M.I.S. for input $G$.

\begin{thm} (Theorem 6.7 and the discussion which follows it, Pages
139-140, in \cite{gj})
\newline
(a) If P $\ne$ NP, then no polynomial time approximation algorithm $A$
for the M.I.S. problem can guarantee
$\vert A(G) - OPT(G) \vert \le k$ for a fixed non-negative constant $k$.

(b) For the M.I.S. problem, for all constants $k$ and $\epsilon > 0$,
no polynomial time approximation algorithm $A$ for M.I.S. can guarantee
$\vert A(G) - OPT(G) \vert \le k \cdot OPT(G)^{(1 - \epsilon)} $ for a
fixed non-negative constant $k$.
$\hfill \Box$
\end{thm}

Let 
$R_A^{\infty}$ be the asymptotic approximation ratio of algorithm $A$
(defined in Page 128 of \cite{gj}).

\begin{thm} (Theorem 6.12, Page 146, in \cite{gj})
\newline
Either the M.I.S. problem can be solved
with a polynomial time approximation scheme, or else there is no 
polynomial time approximation algorithm $A$ for M.I.S. that satisfies
$R_A^{\infty} < \infty$.
$\hfill \Box$
\end{thm}

\subsection{Solvers}\label{sec:solvers}

To execute Linear and Integer Programs, we used GLPK Version 5.0 and
Gurobi 9.5.1 in a
Linux environment. Further information on GLPK is available at:
\tt{https://www.gnu.org/software/glpk/}

Further information on Gurobi can be found at: 
\tt{https://www.gurobi.com}

For non-linear continuous optimisation (such as convex programming), we
used the MINOS solver (coded in AMPL), available through 
\tt{https://www.neos-server.org}.

\section{Binary search}\label{sec:binarySearch}

Let us define an \it{integer solution} to be one where the value of every
$F[j]$ (1 $\le j \le N$) is integer; that is, $F[j] \in \{0, 1\}$.

For each value $Z$ of the objective function checked within the Binary
Search framework, we run an optimisation algorithm once, attempting to
find an integer solution.

In general, one could perform a {binary search} on the objective
function value $Z$, once we determine an upper bound
and a lower bound for $Z$ (obviously, \bf{only} for integer values of
$Z$).

\it{Weakly polynomial algorithm}:
 Since $0 \le UB \le N$, we can say that
$\log (UB)$ is a strongly polynomial term. The algorithm is weakly
polynomial if we run Linear Programming multiple times. There are
weakly polynomial algorithms for LP, but no strongly polynomial algorithm
is known.

Binary search is carried out between two markers, $top$ and $bottom$.  We
fix $top$ to be $Z$'s upper bound (= 20, in the above example for the
1ZC-128 problem).  If there is a feasible integer solution to the Linear
Program where ``($Z = down$)" is another extra constraint, we try the next instance by increasing the
value of $down$; otherwise, we try by decreasing the value
of $down$.  (We want to find the maximum value of $down$ where the
Linear Program optimal solution is feasible integer.)

If there is a M.I.S. of size $k$, there is certainly a M.I.S. of size
$k-1$; just remove one of the vertices in the solution set. From this, it
follows that if there exists a feasible independent set for a particular
value of $down$, say, $down_A$, there is a feasible independent set for
\it{every} value of $down$ that is $\le$ $down_A$.
Hence, once we find an integer optimal solution to the linear program
with ($Z = down_A$), we continue our binary search \it{above} by setting
$Z > down_A$, \it{not} below $down_A$.

As mentioned earlier, we include all Edge Inequalities from Problem 1
(Page \pageref{original_IP}) in every instance that we solve. ~Hence
during Binary Search, for values of $Z$ higher than $OPT (I.P.)$, it is
impossible to find an integer feasible solution.

\subsection{Alternate optimal solutions at extreme points}\label{sec:OEP}

\begin{rem}
For Integer Programs in general, an optimal integer solution could lie in
the \it{interior} of the feasible region of the Linear Programming
relaxation (LP relaxation).  However, for 0-1 (or binary) integer
programs such as M.I.S, optimal integer solutions can \it{always} be
found in one of the extreme points of the LP relxation.  In fact, for
binary integer programs, \bf{every} feasible integer solution will be at
an extreme point of the LP relaxation.
$\hfill \Box$
\end{rem}

It is easy to show the above. For any solution $x$ where every component
$x_i$ is either 0 or 1, $x_i$ cannot be written as a convex combination
of two values $y_i$ and $z_i$ where both are in the feasible range $0 \le
y_i, z_i \le 1$ \bf{and} at least one of them is different from $x_i$.
(If $y_i > 0$, then $z_i < 0$ which is disallowed.)

\begin{rem}
\bf{Open Problem}: Since feasible integer solutions are guaranteed to be
located at extreme points of the LP relaxation, does it make 0-1 Integer
Programs easier to solve than general integer programs? For example, can
we reach an optimal integer solution by doing a ``small" number of
Simplex pivots from an optimal solution to the LP relaxation?
$\hfill \Box$
\label{rem:binaryIP}
\end{rem}

\begin{equation}
\begin{array}{rcl}
\mbox{Maximise} ~  Z_3 &  = & \sum_{j \in V} F[j] \\ 
\mbox{Subject to} & & \\  [2mm]
F[i] + F[j] & \le & 1 ~~ \forall ~ (i,j) \in E \\ [2mm]
\sum_{j \in V} F[j] & = & k ~~~ (k ~ \mbox{is a known value}) \\ [3mm]
 0 \le F[j] \le 1 &  &  \forall ~ j \in V.
\end{array}
\label{eq:binarySearchLP}
\end{equation}

\section{Convex Programming and a bin-packing type of approach}

In Convex programming (or convex optimisation), we minimise a convex
objective function (or maximise a concave objective function) over a
convex feasible region.
For convex optimisation problems, a local optimal solution is also
globally optimal. 

From the experiments in previous sections, it is clear that we need to
search through a large number of extreme points before reaching an
integer solution to Problem (\ref{eq:binarySearchLP}) in Remark
\ref{rem:binaryIP} (Page \pageref{eq:binarySearchLP}). To avoid this, let
us attempt a convex optimisation approach; that is, minimising a convex
objective function within a convex space.

Actually the feasible space is bound by linear constraints since we will
use exactly the same constraints as in Problem (\ref{eq:binarySearchLP}).

Consider the constraint 
\begin{equation}
\sum_{j \in V} F[j] = k
\end{equation}
where $k$ is a positive integer.  For example, consider $k = 6$ and $N =
20$.

We can think of this as having to place 6 ``unit-size" items into 20
``unit-size" bins.  The 6 unit-size items could be broken into smaller
items --- for example, into 12 items of size 0.5 each (but this would
require 12 bins).

The sum of the sizes of all items should be equal to 6 (or, $k$, in
general). Each bin is allowed to hold only one item.  The number of
available bins is 20 (or, $N$ in general).

Another example could be breaking the 6 unit-size items into 13 items of
sizes 0.3, 0.4, 0.2, 1.0, 1.0, 0.5, 0.25, 0.25, 0.25, 0.75, 0.35, 0.65
and 0.1, for which we need 13 bins (each bin is allowed to hold only one
item, however small or large).

The goal is to minimise the number of used bins.  The optimal
way of doing this is to fill $k$ (= 6) bins to make them completely full
(i.e., unit-sized items into unit-sized bins), and leave the other $N-k$
(= 20$-$6 = 14) bins empty. 

For this example, there are multiple optimal solutions, such as (a) Bin
numbers 1, 2, 3, 4, 5 and 6; (b) Bin numbers 1, 3, 5, 6, 11 and 12, etc.
As long as~ $k < N$, there will be multiple optimal solutions.

\subsection{Convex objective function}

So then, our objective is to develop a convex objective function which we
would like to minimise, subject to the constraints in Problem
(\ref{eq:binarySearchLP}) in Remark \ref{rem:binaryIP}.. Some salient
points:

\benum
\item
We would like to minimise the number of used bins. 
\item
Hence the larger the size of an item in a bin $j$, the lower should be
the cost of leasing this bin.  Accordingly, empty bins will be charged
the highest leasing cost and full bins will be charged the lowest.
\item
However, at the same time, since the cost of leasing an empty bin is
high, we cannot afford to have too many empty bins.
\eenum

Let $x_j (= F[j])$ be the size of the item in Bin $j$.  Since we have $N$
bins, and the sum of the sizes of the items over all bins is $k$, we have
the following constraint:
\begin{equation}
\sum_{j \in V} x_j = k,
\label{eq:itemSizeSum}
\end{equation}
where $V$ is the set of all bins and $|V| = N$.

Let $f(x_j)$ (0 $\le x_j \le$ 1) be the convex cost function that we
need.

\subsection{Breakup scenarios}\label{sec:breakupScenarios}

Let us consider several scenarios where we break up a unit-sized item
into several items of smaller sizes.  This may not get us a ``perfect"
convex function $f(x)$ that we are looking for, but it should get us
close enough.

\benum
\item
Break a unit-sized item into 20 items, each of size 0.05 (that is, 1 =
0.05 * 20):

\it{This case is needed only when $N \ge$ 20.}

Case 1 (20 bins, each with an item of size 0.05): Total Cost =
20$f$(0.05).

Case 2 (one bin full with the unit-sized item, the other 19 bins are
empty): Total Cost: $f$(1) + 19$f$(0).

We require that Case 2 have a lower cost than Case 1:
\newline
$f$(1) + 19$f$(0) $\le$ 20$f$(0.05).

\item
Break a unit-sized item into 10 items, each of size 0.1 (that is, 1 =
0.1 * 10):

\it{This case is needed only when $N \ge$ 10.}

Case 1 (10 bins, each with an item of size 0.1): Total Cost =
10$f$(0.1).

Case 2 (one bin full with the unit-sized item, the other 9 bins are
empty): Total Cost: $f$(1) + 9$f$(0).

We require:~ $f$(1) + 9$f$(0) $\le$ 10$f$(0.1).

\item
1 = 0.2 * 5:

Case 1 Total Cost = 5$f$(0.2).

Case 2 Total Cost = $f$(1) + 4$f$(0).

We require:~ $f$(1) + 4$f$(0) $\le$ 5$f$(0.2).

\item
1 = 0.25 * 4:

Case 1 Total Cost = 4$f$(0.25).

Case 2 Total Cost = $f$(1) + 3$f$(0).

Requirement:~ $f$(1) + 3$f$(0) $\le$ 4$f$(0.25).

\item
1 = 0.3333 * 3:

Case 1 Total Cost = 3$f$(0.3333).

Case 2 Total Cost = $f$(1) + 2$f$(0).

Requirement:~ $f$(1) + 2$f$(0) $\le$ 3$f$(0.3333).

\item
1 = 0.001 * 1000 (needed only when $N \ge$ 1000):

Case 1 Total Cost = 1000$f$(0.001).

Case 2 Total Cost = $f$(1) + 999$f$(0).

Requirement:~ $f$(1) + 999$f$(0) $\le$ 1000$f$(0.001).

\item
Two items, each of size 0.5 (that is, 1 = 0.5 + 0.5):

Case 1 Total Cost = 
2$f$(0.5).

Case 2 Total Cost = 
 $f$(0) + $f$(1).

Requirement:~ $f$(0) + $f$(1) $\le$ 2$f$(0.5).

\item
(Two items) 1 = 0.001 + 0.999:

Case 1 Total Cost = 
$f$(0.001) + $f$(0.999).

Case 2 Total Cost = 
$f$(0) + $f$(1).

Requirement:~ $f$(0) + $f$(1) $\le$ $f$(0.001) + $f$(0.999).

\item
(Two items) 1 = 0.02 + 0.98:

Case 1 Total Cost = 
$f$(0.02) + $f$(0.98).

Case 2 Total Cost = 
$f$(0) + $f$(1).

Requirement:~ $f$(0) + $f$(1) $\le$ $f$(0.02) + $f$(0.98).

\item
1 = 0.05 + 0.95:

Case 1 Total Cost = 
$f$(0.05) + $f$(0.95).

Case 2 Total Cost = 
$f$(0) + $f$(1).

Requirement:~ $f$(0) + $f$(1) $\le$ $f$(0.05) + $f$(0.95).

\item
1 = 0.15 + 0.85:

Case 1 Total Cost = 
$f$(0.15) + $f$(0.85).

Case 2 Total Cost = 
$f$(0) + $f$(1).

Requirement:~ $f$(0) + $f$(1) $\le$ $f$(0.15) + $f$(0.85).

\item
1 = 0.3 + 0.7:

Case 1 Total Cost = 
$f$(0.3) + $f$(0.7).

Case 2 Total Cost = 
$f$(0) + $f$(1).

Requirement:~ $f$(0) + $f$(1) $\le$ $f$(0.3) + $f$(0.7).

\eenum

The goal then is to find a convex function $f(x)$ (0 $\le x \le$ 1) that
satisfies the requirements of each of the scenarios above.  The modified
version of Problem (\ref{eq:binarySearchLP}) is as follows:
\begin{equation}
\begin{array}{rcl}
\mbox{Minimise} ~  Z_4 &  = & \sum_{j \in V} f(x_j) \\ 
\mbox{Subject to} & & \\  [2mm]
x_i + x_j & \le & 1 ~~ \forall ~ (i,j) \in E \\ [2mm]
\sum_{j \in V} x_j & = & k ~~ (k ~ \mbox{is a known value}) \\ [3mm]
 0 \le x_j \le 1 &  &  \forall ~ j \in V.
\end{array}
\label{eq:binarySearchLPconvex}
\end{equation}

It is known that the sum of convex functions is also a convex function.
Hence the summation ~$\textstyle \sum_{j \in V} f(x_j)$~ is also convex.
Observe that the objective has been changed from \it{maximise} to
\it{minimise}.

\subsection{Experiments to find convex functions}\label{sec:exptFindCvxFunc}

As said earlier, the 12 breakup scenarios above may not get us a
perfect convex function $f(x)$ that we are looking for, but it should
get us close enough.

To minimise zero errors in numerical computation, we add a requirement
that each of the $N$ bins should contain an item of length at least $w$,
where $0 \le w \le 1$.  To accommodate this new requirement, we modify
the constraint $\textstyle \sum_{j \in V} x_j = k$ in
(\ref{eq:binarySearchLPconvex}) to be as follows:
\begin{equation}
\sum_{j \in V} x_j  =  k  + (N-k)w. \\ [3mm]
\end{equation}
A modified form of (\ref{eq:binarySearchLPconvex}) would be:

\bf{Problem 3}.
\begin{equation}
\begin{array}{rcl}
\mbox{Minimise} ~  Z_5 &  = & \sum_{j \in V} f(x_j) \\ 
\mbox{Subject to} & & \\  [2mm]
x_i + x_j & \le & 1 + w, ~~ \forall ~ (i,j) \in E \\ [2mm]
\sum_{j \in V} x_j & = & k + (N-k)w ~~~ (k ~ \mbox{is a known value}) \\ [3mm]
 w \le x_j \le 1 &  &  \forall ~ j \in V.
\end{array}
\label{eq:binarySearchLPwithW}
\end{equation}

In order to find an appropriate $f(x)$, we ran experiments.  We tested
functions of the form
\begin{equation}
f(x) = \left[ t + M x + r (1 - x) + \frac{y} {x + s} \right]^p
~~~ (0 \le x \le 1, ~ p \ge 1)
\label{eq:convexFunctionA}
\end{equation}
with different values for parameters $p$, $t$, $K$, $r$, $s$, $w$ and $y$.

For instance, in one experiment, we tried the following values:

\medskip

\begin{rem}
\textsf{
\hspace*{8mm}
    for (p=1.0; p $<=$ 2.0; p = p + 0.2)
       \newline \hspace*{16mm}  
       for ($i_1$ = 0; $i_1$ $<=$ 25; $i_1$++) \{
       \newline \hspace*{24mm}  
           t = 0.000001 $(-4)^{i_1}$;
       \newline \hspace*{24mm}  
           for ($i_2$ = 0; $i_2$ $<=$ 25; $i_2$++) \{
       \newline \hspace*{32mm}  
                M = 0.000001 $(-4)^{i_2}$;
       \newline \hspace*{32mm}  
                for ($i_3$ = 0; $i_3$ $<=$ 25; $i_3$++) \{
       \newline \hspace*{40mm}  
                    r = 0.000001 $(-4)^{i_3}$;
       \newline \hspace*{40mm}  
                    for ($i_4$ = 0; $i_4$ $<=$ 25; $i_4$++) \{
       \newline \hspace*{48mm}  
                        s = 0.000001 $(-4)^{i_4}$;
       \newline \hspace*{48mm}  
                        for ($i_5$ = 0; $i_5$ $<=$ 24; $i_5$++) \{
       \newline \hspace*{56mm}  
                            w = 0.0000001 $(2^{i_5})$;
       \newline \hspace*{56mm}  
                            for ($i_6$ = 0; $i_6$ $<=$ 25; $i_6$++) \{
       \newline \hspace*{64mm}  
                                 y = 0.000001 $(-4)^{i_6}$;
}
\label{rem:forLoops}
$\hfill \Box$
\end{rem}

Apart from $w$ (which represents the minimum item size in a bin, and
hence we need $w$ to be $\ge$ 0), all other parameters are allowed to be
positive or negative.

\medskip

We found different convex functions for several 
($p$, $t$, $K$, $r$, $s$, $w$, $y$) combinations
which satisfied the requirements for
each of the 12 breakup scenarios of Section \ref{sec:breakupScenarios}.

\ul{How to check whether a function is convex?}:
This is explained within the
file \tt{convexity-check.h} (see Sec. \ref{sec:softwareURL} for the URL).
~ Rather than use a strictly binary convex or non-convex distinction, we
use a \it{degree of convexity} or a \it{convexity measure}.  We do this
by checking whether the function is convex within small sub-intervals.
This is explained in the file \tt{convexity.h}.

When $s < 0$, of course, there will be singularity (i.e., function
approaching extreme values) when ($x + s$) is close to zero.  But in this
region, we can interpolate the function between two points $x_1$ ($< -s$)
and $x_2$ ($> -s$) that are on either side of $(-s)$.

\subsection{Empty bins}\label{sec:emptyBins}

A key difficulty we first experienced with our experiments had to do with
empty bins.  For example, consider the case when $N$ = 18, $k$ = 6 and a
convex function $f(x_i) = (x_i-1)^2$, $0 \le x_i \le 1$.

We have 18 bins available, each with a capacity of one.
~For an empty bin $i$, the cost is the highest; $f(x_i) = f(0) = 1$.

If the 6 (unit-sized) items are placed in their own bins, without
breaking any item into smaller pieces, then 6 bins are fully filled and
the remaining 12 are empty.  The total cost is $Z_1$ = 6$f(1)$ + 12$f(0)$
= 6(0) + 12(1) = 12 units.

However, if each item is broken into 2 pieces of size (1/2) each, for
each of the 12 occupied bins, $x_i$ = 1/2. The other 6 bins will be
empty. ~The total cost is $Z_2$ = 
12$f(1/2)$ + 6$f(0)$ = 12$(1/2 - 1/)^2$ + 6(1) = 12(1/4) + 6 = 9 units.

Since $Z_2 < Z_1$, the second (non-integer) solution is closer to optimal
than the first; hence we fail to obtain an integer solution for this
instance if we minimise the convex function $f(x_i) = (x_i-1)^2$.

To overcome this difficulty, we introduce another condition as below, in
addition to the 12 conditions (scenarios) we considered in Sec.
\ref{sec:breakupScenarios}:

\begin{equation}
k f(1) + (N-k)f(0) < (2k)f(1/2) + (N-2k)f(0).
\end{equation}

Applying these 13 conditions, were we able to find convex functions which
when minimised (for a given $N$ and $k$), produced 0-1 integer
optimal solutions? That is, optimal $x_i \in \{0, 1\}$ for every bin
$i$ where 1 $\le i \le N$?

The answer is yes, for some small instances such as (a) $N$ = 18 and $k$
= 6; ~ (b) $N$ = 18 and $k$ = 8, ~using the MINOS solver at
\tt{www.neos-server.org}.

\begin{rem}
(a) For the MIS model solved using the MINOS solver, all constraints are
linear.  Hence the feasible region is a convex space.  (However, the
function minimised is \bf{not} linear).

(b) From MINOS documentation, we understand that it produces locally
optimal solutions. If the function minimised is convex, we know that a
locally optimal solution is also globally optimal.
$\hfill \Box$
\end{rem}

Here is one such convex function\footnote{\ul{Note}: For several convex
functions, after roughly 600 iterations in each case, the
MINOS solver returned an optimal solution; however it described the
solutions as ``either infeasible or a bad starting basis"! ~But when we
tested these solutions with an Integer Programming solver such as GLPK or
GUROBI, they turned out to be perfectly feasible and optimal for the
I.P. ~So I have NO idea why MINOS would describe these solutions as
infeasible when they are the optimal solutions that we are looking for.}
 that returned an optimal solution
 for $N$ = 18 and $k$ = 6 with parameters as described in Eq.
(\ref{eq:binarySearchLPconvex}):

Degree of convexity= 991/1000, ~$p$= 2, ~$t$= 268.435456,
~$M$= 1.048576, ~$r$= 268.435456, ~$s$= 0.000001, ~$y= -$0.000064,
and $w$= 0.000001.

For $N$ = 18 and $k$ = 8, here is one such convex function that returned
an optimal solution:

Degree of convexity= 992/1000, ~$p$= 2, ~$t$= 4294.967296, 
~$M$= 16.777216, ~$r$= 268.435456, ~$s$= 0.000001, ~$y= -$0.000004,
and $w$= 0.

Experiments are ongoing to find general convex functions for higher
values of $N$ that pack the $k$ items into a minimum number of bins (that
is, into $k$ bins).

We should do \bf{Curve Fitting} to generate smooth convex functions instead
of using functions such as (\ref{eq:convexFunctionA}) directly in convex
optimisation solvers.

When $(k/N)$ is close to either one or zero, the Linear Program in
(\ref{eq:binarySearchLP}) in Page \pageref{eq:binarySearchLP}
can be solved to optimality fast through other methods.  Other
cases will be (much) more challenging.  In many (or perhaps even most)
instances, when $k \approx N$, optimal solutions to problems in
(\ref{eq:binarySearchLP}) and (\ref{eq:binarySearchLPconvex}) will
\it{not} be integer.

\subsubsection{Modelling}

For modelling the objective function and constraints, I suggest
following Disciplined Convex Programming (DCP). ~Please refer to 
\texttt{http://dcp.stanford.edu}.  After I implemented a few DCP rules,
the performance of the MINOS solver significantly improved (finding an
optimal solution in 118 iterations instead of 600).

The changes I implemented were:

(a) Instead of writing the constraint $~G_i - \frac{y}{x_i+s} = 0$, ~I
expressed it as $~G_i(x_i+s) - y = 0$;

and

(b) Added a constraint explicitly stating that:
\newline
\hspace*{10mm} $ t + M x_i + r (1 - x_i) + G_i \ge 0$
~ (in addition to the constraint $~G_i(x_i+s) - y = 0$). 

\subsection{Suggestions for further research along these lines}\label{sec:future}

When $s < 0$, rather than doing a ``curve fitting" between $(-s -
\epsilon)$ and $(-s + \epsilon)$ (Sec. \ref{sec:exptFindCvxFunc}), we
could let each $x_i$ belong to the interval [$-s + \epsilon$, 1].
This means, we can avoid the non-convexity when $x < -s + \epsilon$.

If we let $x_i \ge -s + \epsilon$, then in order to get the solution we
desire (that is, $k$ bins filled to capacity), we should require the
remaining ($N-k$) bins to contain an item of size ($-s + \epsilon$). Then
condition (\ref{eq:itemSizeSum}) will now become:
\begin{equation}
\sum_{j \in V} x_j = k + (N-k)(-s + \epsilon).
\label{eq:itemSizeSum2nd}
\end{equation}

\subsubsection{Software Webpage URL}\label{sec:softwareURL}
 
The software programs are available at the following URL:

\tt{https://sites.google.com/view/all-optimisation-slides/}

The files relevant to the convex programming of Section
\ref{sec:exptFindCvxFunc} are \texttt{bin-packing-v2.c},
\newline
\texttt{bin-packing-v4.c} and \tt{convexity-check.h}.

\subsection{A polynomial convex function}\label{sec:polyConvexFunc}

For at least one tuple of ($N$, $k$) values, we were able to successfully
find a convex function that is a polynomial.

Also note that once it works for a single ($N$, $k$) pair, it applies to
\bf{all} M.I.S. instances with the same number $N$ of vertices and optimal
(integer) solution value $k$ (and perhaps for other, non-optimal integer
values such as $k \pm i$).

So next, we tried a polynomial function of the form:
\begin{equation}
f(x) = a_4x^4 + a_3x^3 + a_2x^2 + a_1x + C + \frac{b_1}{x} + 
       \frac{b_2}{x^2} + \frac{b_3}{x^3} + \frac{b_4}{x^4} ~~~
    (0 \le x \le 1).
\label{eq:polyCvxFunc}
\end{equation}

Thus we need to find 10 parameters: 
$a_4, ~ a_3, ~ a_2, ~ a_1, ~ C, ~ b_1, ~ b_2, ~ b_3, ~ b_4$ and $w$.

\begin{rem}
Algorithms that determine values for the 10 parameters need
\bf{not} run in polynomial time.  This is because, once determined, these
values can be used to solve a large range of M.I.S. instances.
$\hfill \Box$
\end{rem}

Previously, we used an exhaustive manual search to determine the
parameters (as in Remark \ref{rem:forLoops}).  Now, let us try to obtain
them using optimisation.

When we tried to determine the above (values for the 10 parameters), we
found that one function that needed to evaluated multiple times was of
degree 4 in the variable $w$, which was the following (we call this
$funcW$):
\begin{equation}
funcW = a_4w^4 + a_3w^3 + a_2w^2 + a_1w + C + \frac{b_1}{w} + 
       \frac{b_2}{w^2} + \frac{b_3}{w^3} + \frac{b_4}{w^4} ~~~
    (0 \le w \le 1).
\label{eq:funcWDef}
\end{equation}

Besides $w$, the other 9 entities 
$a_4, ~ a_3, ~ a_2, ~ a_1, ~ C, ~ b_1, ~ b_2, ~ b_3$ and $b_4$ are also
unknowns whose values we need to determine using optimisation, which, as
it stands currently, is non-linear.

 ~However, if we fix $w$, the optimisation becomes linear,
and we can obtain values for the other 9 variables by solving a Linear
Program (LP).

So that is what we did; we tried various values for $w$, and each time,
the LP returned values for the other 9 variables. Note that NO objective
function is necessary; we only need a feasible solution to the LP (you
can add an objective function if you wish).

The LP is included in Appendix A in AMPL format.

\begin{rem}
\bf{Connected graphs}. ~For connected graphs, the maximum size of
independent sets is at most $\lceil N/2 \rceil$.  So it is sufficient to
do the binary search for values of $k$ that are $\le \lceil N/2 \rceil$. 
$\hfill \Box$
\end{rem}

\subsubsection{Verification that the function generated is indeed convex}

How to verify that the function  (\ref{eq:polyCvxFunc}) generated is
convex?  For this, we ensure that the second derivative is positive at
several points; in particular, at $x$ = 0.0, 0.00001, 0.00002, 0.00003,
0.00004, $\cdots$, 1.0. (But we verify this only for $x \ge w$.  For $x <
w$, it is \bf{not} necessary since the size of the item in every bin is
required to be at least $w$.)

The second derivative of (\ref{eq:polyCvxFunc}), of course, is given by:
\begin{equation}
f^{\prime \prime} (x) = 12 a_4x^2 + 6 a_3x + 2 a_2 + 2 \frac{b_1}{x^3} + 
     6 \frac{b_2}{x^4} + 12 \frac{b_3}{x^5} + 20 \frac{b_4}{x^6} ~~~
    (0 \le x \le 1).
\label{eq:secondDeriv}
\end{equation}

\begin{rem}
In this section, we solve every instance of Problem
\ref{eq:binarySearchLP} in two steps:
\begin{itemize}
\item
Step A:  Solve an LP to determine the 9 parameters $C$, $a_i$ and $b_i$,
1 $\le i \le$ 4 (we usually used the Gurobi solver in this step); and
\item
Step B:  Solve a non-linear continuous optimisation version of Problem
\ref{eq:binarySearchLPconvex} (similar to the one in Appendix B), with
linear constraints and a non-linear function $f(x_j)$ (we usually solved
this with the MINOS solver).
\end{itemize}
Of course, for Step B, we try to borrow parameters from other instances
rather than generating parameters for every instance.

NO ``initial (starting) solution" was provided to the models in these
experiments.
$\hfill \Box$
\label{rem:twoStepProc}
\end{rem}

\subsubsection{For the case when N = 25 and k = 4}\label{sec:convexPolyInst}

The convex program that we used is in Appendix B in AMPL format.  

(\it{Step A}) 
For $N$ (number of vertices) = 25 and $k$ (optimal M.I.S. solution value)
= 4, by setting $w$ = 0.005, we obtained the following values for the 9
other parameters from the LP:

\hspace*{10mm}
$C$ = 50014.9;

\hspace*{10mm}
$a_1$ = 9999950000; ~~ 
$a_2$ = 1.60683; ~~ 
$a_3$ = $-$1.92801; ~~ 
$a_4$ = 0.876627;

\hspace*{10mm}
$b_1$ = $-$0.00314755; ~~ 
$b_2$ = 8.11163e$-$05; ~~ 
$b_3$ = $-$5.21308e$-$07; ~~ 
$b_4$ = 1.10935e$-$09.

Note: Since we are adding an item of size $w$ to the remaining ($N-k$)
bins, the classical edge inequality ``$X[x] + X[y] \le 1$" for edge 
($x$, $y$) should be modified to ``$X[x] + X[y] \le 1 + w$".

(\it{Step B}) 
These 9 values (together with $w$ = 0.005) are then used in functions
(\ref{eq:polyCvxFunc}) and (\ref{eq:funcWDef}), as well in the convex
program (\ref{eq:binarySearchLPconvex}) in Page
\pageref{eq:binarySearchLPconvex}.
We then solved 
(\ref{eq:binarySearchLPconvex}) using the solver MINOS, which returned
the following values for $x_j$:

\begin{center}
\begin{tabular}{|ll|ll|ll|ll|ll|ll|}
\hline
$j$ & $x_j$ & $j$ & $x_j$ & $j$ & $x_j$ & $j$ & $x_j$ & $j$ & $x_j$ & $j$
& $x_j$ \\ 
\hline
\hline
1 & 0.005 &  5 & 0.005 &  9 & 0.005 & 13 & 0.005 & 17 & 0.005 & 21 & 0.005 \\
\hline
2 & 0.005 &  6 & 0.005 & 10 & 0.005 & 14 & \bf{1} & 18 & 0.005 & 22 & 0.005 \\
\hline
3 & 0.005 &  7 & 0.005 & 11 & 0.005 & 15 & 0.005 & 19 & 0.005 & 23 & \bf{1} \\
\hline
4 & 0.005 &  8 & 0.005 & 12 & 0.005 & 16 & 0.005 & 20 & 0.005 & 24 & \bf{1} \\
\hline
25 & \bf{1} &   &   &  &  & \multicolumn{6}{|c|}{Number of MINOS
iterations = 70}  \\
\hline
\end{tabular}
\end{center}

From this, we infer that an optimal solution to the given M.I.S. instance
is the set of vertices $\{$14, 23, 24, 25$\}$, since the corresponding
function values in the above table are equal to 1.0.

Note: For this particular MIS instance, the optimal solution value is 4.
So if we run the LP with $k = 5$ to determine values for the 9 variables
($a_4$, $\cdots$, $b_4$), and if these values are used in the convex
program (\ref{eq:binarySearchLPconvex}), we can never obtain an integer
solution for this MIS instance.

Note that we used ``scale\_option = 2" with the MINOS solver.

\subsubsection{Tighter constraints}\label{sec:tighterConstraints}

The next thing we did was to tighten some of the constraints, to consider
the fact that $w$ may \bf{not} be zero. 

So, if a unit-sized item is split into two items of sizes 0.3 and 0.7 and
placed in two separate bins, then the 2 cases we should compare could be
as follows:
\begin{enumerate}
\item[Case (a)]:
The first bin contains a unit-sized item, and the second bin
contains an item of size $w$.. Hence the total size of the 2 items is (1
+ $w$).  ~ Total cost = $f$(1) + $f(w)$.

\item[Case (b)]: The total size (1 + $w$) is split into 2 bins as follows.
The first bin contains an item of size (0.3 + $w$/2), and the second 
bin contains an item of size (0.7 + $w$/2). 
~ Total cost = $f(0.3 + 0.5w)$ + $f(0.7 + 0.5w)$.

Obviously we want Case (a) to be the winner (with lower total cost than
Case (b)).

Consider another example:  A unit-sized item is split into 10 items of
size 0.1 each.  The two cases to be compared are:

\item[Case (c)]: The first bin contains a unit-sized item, and 9 other bins
contain an item of size $w$ each. ~Hence the total size of the 10 items
is (1 + 9$w$). ~ Total cost = $f$(1) + 9$f(w)$.

\item[Case (d)]: The total size (1 + 9$w$) is split equally into 10 bins, with
each bin containing an item of size (1 + 9$w$)/10 = 0.1 + 0.9$w$.
~ Total cost = 10($f(0.1 + 0.9w)$).
\end{enumerate}

Comparing cases (c) and (d), obviously we want Case (c) to have the lower
cost.

The bin-packing LP model with these modified constraints is presented in
Appendix C.

\medskip

\bf{An instance using the new model}:

~For the new LP model, we used the following options/parameters:
\begin{verbatim}
option gurobi_options 'scale=1';

param N := 25;
param k := 4;
param intvl := 100000;
param eps := 20;
param lowCurv := 1500;
param curv_lower_bound := 0.00000001;

# And hence, w = 0.015.
\end{verbatim}

(\it{Step A} of Remark \ref{rem:twoStepProc}) 
For the LP with the above parameters, Gurobi returned the following values:

\hspace*{15mm} $C$ =  100000000.000082492828369; ~~ 
$a_1$ =  $-$0.000082507074949;

\hspace*{15mm} $a_2$ =  0.000000005; and ~ $a_3 = a_4 = b_1 = b_2 = b_3 = b_4 =$ 0.

(\it{Step B}) 
With these parameters, we ran the M.I.S. model (in Appendix B) with
the MINOS solver (with ``scale\_option = 2") and obtained the following
values for $x_j$:

\begin{center}
\begin{tabular}{|ll|ll|ll|ll|ll|ll|ll|}
\hline
$j$ & $x_j$ & $j$ & $x_j$ & $j$ & $x_j$ & $j$ & $x_j$ & $j$ & $x_j$ & $j$ & $x_j$ & $j$ & $x_j$ \\ 
\hline
\hline
1 & 0.015 &  5 & 0.015 &  9 & 0.015 & 13 & 0.015 & 17 & 0.015 & 21 & 0.015 & 25 & 0.015 \\
\hline
2 & 0.015 &  6 & 0.015 & 10 & 0.015 & 14 & 0.015 & 18 & 0.015 & 22 & \bf{1} & & \\
\hline
3 & \bf{1} &  7 & \bf{1} & 11 & 0.015 & 15 & 0.015 & 19 & 0.015 & 23 & 0.015 & & \\
\hline
4 & \bf{1} &  8 & 0.015 & 12 & 0.015 & 16 & 0.015 & 20 & 0.015 & 24 & 0.015 & & \\
\hline
\multicolumn{14}{|l|}{Number of MINOS iterations = 9}  \\
\hline
\end{tabular}
\end{center}

From the above, we infer that the optimal solution is the set of vertices
\{3, 4, 7, 22\}.

In fact, for the MIS program under MINOS, the values of the 9 parameters
($C$, $a_i$, $b_j$) given above also work for several values of $w$
between 0.0002863 and 0.29989985, sometimes returning a different optimal
solution \{3, 7, 15, 22\}.

Notice the \bf{drastic drop} in the number of iterations from 70 to 9.

\subsubsection{An instance with 150 vertices}\label{sec:N150instance}

We were able to successfully solve an instance with $N$ = 150 (150
vertices), with $k$ = 20.  ~For this, we only needed to add the following
2 conditions as constraints to the model in Appendix C (which is an
addendum to the model in Appendix A):
\begin{equation}
\begin{array}{l}
\mbox{Cost[a unit-sized item in its own bin] + 7*Cost[an item of size $w$ in
its own bin]} \\
 < \mbox{ (8 times the cost of placing an item of size (1 + 
7$w$)/8 in its own bin) + eps.}
\end{array}
\end{equation}
\begin{equation}
\begin{array}{l}
\mbox{Cost[a unit-sized item in its own bin] + Cost[an item of size $w$ in
its own bin]} \\
 < \mbox{ (twice the cost of placing an item of size (0.5 + 0.5$w$) in
its own bin) + 1.5*eps.}
\end{array}
\end{equation}

In the first constraint above, observe that the quantities on both the
right and left sides of the relation symbol ($<$) need 8 bins each.
Similarly in the second constraint, they need 2 bins each.

(\it{Step A} of Remark \ref{rem:twoStepProc}) 
Once we solved the LP model above using Gurobi, with $w$ = 0.008, we
obtained parameter values as in Table \ref{tab:N150parameters}.

\begin{table}[h]
\begin{tabular}{|l|l|l|}
\hline
$C$ = 98999971.4548247457 & 
$a_1$ = 1000028.5881573070 & 
$a_2$ = $-$0.0943214668  \\
\hline
$a_3$ = 0.0617447224 & 
$a_4$ = $-$0.0159219866 & 
$b_1$ = 0.0059862135 \\
\hline
$b_2$ = $-$0.0004836134 & 
$b_3$ = 0.0000142205 & 
$b_4$ = $-$0.0000001472 \\
\hline
\multicolumn{3}{|l|}{~ Ind. Set = \{$4, 14, 22, 24, 37, 63, 64, 78, 91, 94, 95, 116,
126, 131, 139, 141, 142, 143, 146, 150$\}}\\
\hline
\multicolumn{3}{|l|}{ ~ $w$ = 0.008. ~~~ Number of MINOS iterations = 803} \\
\hline
\end{tabular}
\caption{Parameters for a 150 vertex instance ($k$ = 20) and an
independent set.}
\label{tab:N150parameters}
\end{table}

(\it{Step B})
Applying the values in Table \ref{tab:N150parameters} to a non-linear
minimisation problem (similar to the one in Appendix B) with $k$=20, we
obtained a desired result, which is, an ``$x_j$ = 1" value for each of the 20
vertices in an independent set, and ``$x_j$ = $w$ = 0.008" ~for the other
130 vertices.  The independent set obtained is given in the last row of 
Table \ref{tab:N150parameters}.

The Google Drive folder is located here:
\newline
\texttt{https://drive.google.com/drive/folders/1t6aUcniJE0xwGwlBhrzxgzZzZM9QHHlq}

The relevant files are:
\begin{itemize}
\item
\texttt{LP-Model-for-N-150-k-20.txt} (LP model to find the 9 parameters)

\item
\texttt{MIS-N-150-opt-29-k-20.txt} (Non-linear problem to solve MIS) 

\item
\texttt{Results-N-150-opt-29-k-20.txt} 
(Result file, with results from the 2 programs above)
\end{itemize}

\bf{Other values of k}: The parameter values in Table
\ref{tab:N150parameters} work for lower values of $k$ (meaning, $k \le$
20); that is, MINOS returns integer solutions. ~However, for $k \ge 21$,
they \it{do not}. (For the given MIS instance, the optimal integer
solution value is 29.)

\bf{Note}: With the ``scale=2" option, MINOS was able to obtain an
optimal integer solution for $k$ = 20 even when the function $f(x)$ was
non-convex up to $x$ = 0.08.  (Curvature values were negative up to $x$ =
0.08.)

\subsubsection{Experiments for N = 150 and k = 29 with partial initial
solutions}\label{sec:N150k29PartialSoln}

For the instance used in Section \ref{sec:N150instance}:

MINOS appears to have trouble with a good guess for an initial solution.
We obtained good results when we provided partial initial solutions by
specifying values for just five vertices (out of 150).

The partial solutions could be obtained from an output to a polynomial time
heuristic to MIS, for example.

For instance, in one set of experiments, we used the following parameter
values:
$C$ =  -68.5841804663; $a_4$ =  -19.7372961484; $a_3$ =  66.2444754126;
$a_2$ =  -81.978279066; $a_1$ =  3934.9825010133; $b_4$ =  -0.0000013593;
$b_3$ =  0.0004225024; $b_2$ =  -0.0436396992; and
$b_1$ =  1.7970273951.

We used a partial solution $x_5 = x_8 = x_9 = x_{13} = 1$.

Using the above values, for $k = $29 (optimal solution value), MINOS was
able to obtain a complete integer solution (that is, $x_i = 1$ for 29
values of $i$ and $x_i = w$ for the remaining 121 values of $i$) for the
following values of $w$ when we did a neighbourhood search in the
interval $0.015 \le w \le 0.0153$ in steps of $10^{-6}$:

$w$ =  0.015028 (12089 iterations, 0.49 seconds)
\newline
$w$ =  0.015106 (4825 iterations, 0.23 seconds)
\newline
$w$ =  0.015134 (3153 iterations, 0.14 seconds)
\newline
$w$ =  0.015145 (9676 iterations, 0.39 seconds)
\newline
$w$ =  0.015245 (1669 iterations, 0.09 seconds)
\newline
$w$ =  0.015296 (3460 iterations, 0.16 seconds)

\paragraph{Experiments for k = 29, the optimal solution
value.}\label{para:k29Expts}

For the instance used in Section \ref{sec:N150instance}:

Unlike previous experiments, we did \it{not} use constraints that enforce
convexity of the function $f(x)$ in (\ref{eq:polyCvxFunc}).  We solely
relied on making the cost of the integer feasible solution significantly
lower than the cost of other feasible non-integer solutions.

As before, we ran an LP model in Gurobi with $w = 0.015$, but used a
higher $k$ value ($k = 32$).  We obtained the following values for the
parameters:

$C =       29906106.9514391012$, 
\newline
$a_4 =       -1264100.9309055060$, 
\newline
$a_3 =        1719464.8249800189$, 
\newline
$a_2 =        -370171.3208458638$, 
\newline
$a_1 =      -29887909.5760779195$, 
\newline
$b_4 =              0.0106982338$, 
\newline
$b_3 =             -1.7969438713$, 
\newline
$b_2 =            114.3168559025$, and 
\newline
$b_1 =          -3502.4792000884$.

\bf{Partial Solutions}.  Again, when we used a partial solution $x_5 = x_8
= x_9 = x_{13} = x_{16} = 1$, the frequency of complete integer solutions
with the parameters above increases significantly.
  While searching in
the interval $0.149 \le w \le 0.01512$, the complete solution (that is,
$x_i^* = 1$ for 29 values of $i$ and $x_i = w$ for the rest) occurs at 
$w =$  0.014910, 0.014943, 0.014967, 0.015019, 0.015037, 0.015055,
0.015056, 0.015061, 0.015065, 0.015066, 0.015071, 0.015088, 0.015089 and
0.015102 (the best computation time being 0.27 seconds and 2.05 seconds
the worst).

\bf{Without Partial Solutions}. 
Empirically, we observed that increasing the value of $C$ (to some degree)
resulted in better integer solutions. (Of course, strictly speaking, a
constant term in the objective function should \it{not} matter in an
optimisation problem.  However, it seems to make a difference in MINOS.)

For $k = 29$, we obtained a feasible integer solution (that is, with an
optimal value of ~$x_j^* = 1$~ for 29 number of $x_j$ variables) to
Problem 3 (Sec.  \ref{sec:exptFindCvxFunc}, Page
\pageref{eq:binarySearchLPwithW}) as follows:

We increased $C$ by 3$\times$10$^7$ (to 59906106.9514391012). 
We were unable to obtain an integer solution at $w = 0.015$;
hence we proceeded to do a small \it{neighbourhood search}.
Starting from 0.015, in steps of 0.00002, using MINOS, we solved an
instance of Step B (Remark \ref{rem:twoStepProc}) for every $w$ until we
reached $w = 0.01574$ (that is, we solved 38 instances of Step B before
obtaining an integer feasible solution to Problem 3).

Total time taken by MINOS to solve the 38 instances = 26.78 seconds.

(The version of MINOS used in these experiments was different from the
version currently available through the NEOS server.)

All relevant files have been combined into one single file called
\newline
~\texttt{All-files-N-150-k-29-w-01574.txt}~ and is available in the
Google drive folder. 

\subsubsection{Experiments with N=150, k=29, partial solutions and a
different function}\label{sec:150VertexNewFunction}

For the instance used in Section \ref{sec:N150instance},
we experimented with a new function:
\begin{equation}
f(x) = a_4 x^4 + a_3 x^3 + a_2 x^2 + a_1 x + C + 
       b_2 x^{1/2} +  b_3  x^{1/3} + b_4 x^{1/4} ~~~
    (0 \le x \le 1).
\label{eq:fracPolyCvxFunc}
\end{equation}

We also changed the way we used the \it{eps} threshold.  Previously we
used the constraint

\hspace*{2cm} (computed solution) $-$ (desired solution) $\ge$ 1 + \it{eps}.

For these experiments, we changed the constraint to:

\hspace*{2cm} (computed solution)/(desired solution) $\ge$ 1 + \it{eps}.

Using Gurobi (scale option = 1), $eps = 100$ and $w = 0.015$ with an
objective to maximise $f(1)$ (with an upper limit of $f(1) \le 10000)$,
we obtained the following parameter values:

$C = 1074640000$,
$a_1 = 6112650000$,
$a_2 = -3352570000$,
$a_3 = 2720420000$,
$a_4 = -1052140000$,
$b_2 = -23925900000$,
$b_3 = 45313200000$, and
$b_4 = -26890300000$.

No convexity requirements were imposed while obtaining a function with
the parameters above.

With the above parameters, we ran MINOS (with scale option = 2), with two
additional constraints

\hspace*{2cm} {$-1 \le$ (total Cost $-$ desired Cost) $\le$ 9,}

so as to constrain the search space of objective function values.  We
used the initial partial solution (set as constraints in the optimisation
model) $x_5 = x_7 = x_8 = x_9 = x_{13} = 1$, and conducting the search in
the interval ~0.149 $\le w \le$ 0.015047~ in steps of $10^{-6}$. 

Complete integer optimal solutions to the MIS instance were obtained as per Table
\ref{tab:newEpsMethodNewFunc}.

\begin{table}
\begin{center}
\begin{tabular}{|c|c|c|c|c|c|}
\hline
$w$  &  Iterations  &  Run time & $w$ & Iterations & Run time \\
     &              & (seconds) &     &        & (seconds) \\ [2mm]
\hline
0.014906 & 11910 & 0.83 &
0.014915  &  2538  &  0.19 \\
0.014917  &  29789 & 2.21 &
0.014928  &  1568 & 0.1 \\
0.014930  &  14184 & 1.03 &
0.014934  &  3031 & 0.21 \\
0.014936  &  34947 & 0.52 & 
0.014953  &  3071 & 0.15 \\
0.014958  &  36777 & 2.66 & 
0.014965  &  5580 & 0.28 \\
0.014967  &  7757 & 0.4 & 
0.014979  &  9589 & 0.67 \\
0.014986  &  3795 & 0.21 & 
0.014988  &  3726 & 0.27 \\
0.014996  &  49423 & 3.58 & 
0.015007  &  4408 & 0.23 \\
0.015014  &  16983 & 1.12 & 
0.015040  &  4763 & 0.25 \\
0.015043  &  13769 & 0.72 & 
0.015048  &  3553  &  0.19 \\
0.015068  &  43709 & 3.09 & 
0.015092  &  2984  &  0.23 \\
0.015124  &  4365  &  0.22 & 
0.015129  &  4352  & 0.23 \\
0.015158  &  36889 & 2.57 & 
0.015185  &  4179  & 0.35 \\
0.015215  & 14137  &  0.75 &  & & \\
\hline
\end{tabular}
\end{center}
\caption{List of $w$ values in [0.149, 0.015047] that returned optimal complete integer solutions}
\label{tab:newEpsMethodNewFunc}
\end{table}

The relevant files uploaded to the ~\tt{Convex-optimisation-method}~
directory are:

\tt{LP-N-150-k-29-new-function-dec-2023.txt} ~~ and

\tt{Dec-2023-MINOS-program-N-150-k-29.txt}.

(You may need to change the file extensions from \tt{.txt} to \tt{.mod})

As can be observed from the table, the first time we obtained a complete
integer solution was  in Run \# 7 of MINOS, at $w$ = 0.014906 (this is
the 7th run of MINOS, since we started with $w$ = 0.014900).

Thus, we have obtained complete integer solutions in MINOS run numbers 7,
18 ($w$ = 0.014917), 31 ($w$ = 0.014930), 37 ($w$ = 0.014936), 59, 68,
etc.

\subsubsection{Instances with 53 vertices}\label{sec:53vertex}

Next, we tested a few instances with 53 vertices, using the
parameters in Table \ref{tab:N150parameters} (but we experimented with
different values of $w$).

\bf{Instance A}:  The optimal solution value to this
MIS instance is 26. ~For 10 $\le k \le$ 26, a value of $w$ = 0.19
\it{works}; meaning, that it produces a feasible integer solution. That
is, it fills $k$ bins with one unit-sized item in each, and fills (53$-k$)
bins with an item of size $w$ in each.. Then for 1 $\le k \le$ 9, a
value of $w$ = 0.188 works (produces the appropriate integer solution).

The number of MINOS iterations is in the 164-180 range.

\bf{Instance B}:  The optimal solution value to this MIS instance is 20.

For $1 \le k \le 10$, $k = 12$ and $k = 20$: ~A value of $w$ = 0.19
works.

For $k = 11$, ~$w = $ 0.189 works.

For 13 $\le k \le$ 18, we used $w =$ 0.035 and it worked.

If $k =$ 19, $w =$ 0.15 was successful.

The number of MINOS iterations is in the 190-256 range.

\bf{Instance C}: All information is within the file 
\texttt{MIS-model-53-vertex-opt-23-k-20.txt}.

See the google drive folder (cf. Sec. \ref{sec:N150instance}) for the
file mentioned above and these 2 files:
\newline
\texttt{53-vertex-k-26-MIS-model-A.txt} ~and~
\texttt{N-53-opt-20-k-11-instance-B.txt}.

The number of MINOS iterations is in the 208-212 range.

For these new LP model files, note that we don't compute the function
values at discrete points any more; instead, we compute function values
at exact points (such as $f(0.1567)$, $f(0.1852)$, etc.) as and when we
need these values.

\begin{rem} 
(a) As observed in these experiments, the ratio ($k/N$) is \bf{not} the
only factor that affects the results.  It depends on the specific
instances as well. 

(b) If we are able to produce an integer solution for a particular value
of $k$, for example $k$ = 20, there is NO need to look for integer
solutions for $k <$ 20; we may only need to find integer solutions for $k >$
20 (by finding sets ($C$, $a_i$, $b_j$, $w$) of parameters that {\em
work} for these $k$ values).

(c) In Step B of Remark \ref{rem:twoStepProc}, while deciding whether a
vertex $j$ should be included in the independent set, it is \bf{not}
necessary that $x_j$ be equal to one, or higher than a certain fixed
threshold such as 0.8 or 0.9.. If $x_j$ is strictly higher than
$(1+w)/2$, vertex $j$ can be included in the independent set.  This is
because, the edge inequality is ~$x_i$ + $x_j$ $\le$ $(1+w)$~ for every
edge $(i,j)$ in the given input graph.

(d) In Appendix B, we strictly need a constraint that states
``{\em \tt{costDifference = 0}}".  However, relaxing this to ~ 
``$-${\em \tt{1000} $\le$ \tt{costDifference} $\le$ \tt{1000}}" ~ 
helps in finding a desired integer solution.

$\hfill \Box$
\label{rem:twoStepPoints}
\end{rem}

\subsubsection{An instance with 256 vertices}\label{sec:oeis_1dc_256}

As we have said before, \bf{transferability} is the key (that is, using
parameters developed for an instance A, for example, to obtain integer
solutions for a different instance B). ~We next focus on the hard MIS
instances posted at \tt{https://oeis.org/A265032/a265032.html}.  

As for the ``1DC-256" instance with 256 vertices and 3839 edges (whose
optimal MIS solution value is 30), using the parameters in Table
\ref{tab:N150parameters} with $w$ = 0.008, we were able to obtain integer
solutions for $k \le$ 22. 

\it{Number of MINOS iterations for the instance above was 872.}

~For higher $k$ values, testing continues.

\subsubsection{Recognising an Independent Set from a Fractional
Solution}\label{sec:recog_integer}

Let's use $x_i$ in place of $F[i]$.

For a solution to be recognised as an ``independent set", it is \bf{not}
necessary that every $x_i$ is a member of the set $\{1, w\}$.  For
example, consider an edge ($a$, $b$).  The edge inequality associated
with this edge is 
\[
x_a + x_b \le (1 + w).
\]

Suppose in a solution $S_1$, let $x_a$ = 0.75(1 + $w$) and 
$x_b  = 0.25(1 + w$).. We say that vertex ~$a$~ \bf{dominates}
the edge ($a$, $b$), and hence, ~$a$~ can be included in the
\it{independent set} of the MIS instance (but \bf{not} vertex ~$b$).

For a different edge adjacent to vertex ~$a$, say edge ($a$, $c$), the
edge inequality is ~ $x_a$ + $x_c$ $\le$ (1 + $w$).
 ~ Since $x_a$ = 0.75(1 + $w$), we require that 
$x_c \le  0.25(1 + w$).~ Again, this time for edge ($a$, $c$), ~$a$~
\it{dominates} ~$c$~ (since $x_a > x_c$).

Thus in this example, vertex ~$a$~ is a member of the independent set, but
vertices ~$b$~ and ~$c$~ are \bf{not}.

In practice, since the RHS of every edge inquality is $(1+w)$, we can
include a vertex ~$t$~ in the \it{independent set}~ if $x_t$ is slightly
higher than ~$0.5(1+w)$.

\subsubsection{Some more constraints}\label{sec:edgeIneq}

The quantity ~$x_i x_j$~ achieves its maximum when ~$x_i$ = $x_j$. ~Since
$x_i + x_j$ $\le$ $(1+w)$ for every edge, the values of $x_i$ and $x_j$
are ~$(1+w)/2$~ each when ~$x_ix_j$~ reaches its maximum.  To avoid
solutions where $x_i$ = $x_j$ = $(1+w)/2$, we can introduce the following
non-linear constraint in Step $B$ (Remark \ref{rem:twoStepProc}) for
every edge in the input graph:

\begin{equation}
x_i  x_j ~ \le ~ \left( \frac{1+w}{2} \right)^2 - \epsilon, ~~~~ \forall ~ (i,j) \in E
\end{equation}
where ~$\epsilon$ ($> 0$)~ is a small positive constant.

Also, given an edge constraint such as $x_i + x_j \le (1+w)$, observe
that ~$w^2 \le x_i x_j \le w$ ~(since $x_i$, $x_j$ $\ge w$).  This gives
us an additional constraint for every edge:
\begin{equation}
x_i  x_j ~ \le ~ w, ~~~~ \forall ~ (i,j) \in E.
\label{eq:XiXj0}
\end{equation}
When $w = 0$, the above inequality becomes the classical constraint ~ $x_i x_j
= 0$.

\subsubsection{Multiple runs within AMPL}\label{sec:multipleAMPL}

For doing multiple runs (with different $w$ values, for example), it is
\it{not} necessary to write a C or Python program.  This can be done
within AMPL.

The following options can be included in AMPL, to make each run fully
independent of previous runs:

\texttt{option reset\_initial\_guesses 1, dual\_initial\_guesses 1,
send\_statuses 0;}

If the options above are \it{not} used, then it is possible that AMPL
will use the solution from the previous run as a starting point for the
current run.

Once the above options are placed at the beginning of the model file, we
can place a ``for" loop after the \it{model} and \it{data} sections of
the file:

\begin{verbatim}
for {T in 0.01 .. 0.05  by 0.00001} {
     let w := T;
     solve;
     display N, eps, k, X;
     printf "status= %i  w= %9.6f \n", solve_result_num, w;
}
\end{verbatim}

(Of course, ~~\texttt{param w;}~~ should be declared before the model and
data sections.)

The loop above runs the model for values of $w$ from 0.01 to 0.05, in
steps of 0.00001.

\subsubsection{Conclusion for this section}

\begin{rem}
\bf{Questions}:

(a) As $N$ grows, what constraints are necessary and/or
sufficient to produce an appropriate set of 10 parameters ($a_1$,
$\cdots$, $b_4$) which would result in a convex optimisation model for
M.I.S. ?

(b) When we try to produce optimal solutions to MIS, why do a particular
set of parameters work for one instance but \bf{not} another, even for
the same values of $N$ and $k$? (as in Sec.  \ref{sec:53vertex})

(c) If $k$ is close to either zero or $\lceil N/2 \rceil$, can the integer
programming version of the Problem in (\ref{eq:binarySearchLPconvex}) in
Sec.  \ref{sec:breakupScenarios} solved to optimality fast using
classical methods such as \underline{branch and bound}?  The problem is
just a re-distribution of the unit-sized items into different bins.

(d) \it{Conjecture}.
From the results file, it appears that some constraints are more critical
than others.  It appears that constraints involving the quantities 
{\em Nkw4Diff, Nkw3Diff, Nkw2Diff, W3, V0199} are more critical than others.
Does this mean that these constraints are sufficient to obtain good
values for the parameters ($C$, $a_i$, $b_j$, $w$)?
$\hfill \Box$
\end{rem}

(The lower the values of Nkw*Diff, W*, V*, the tighter the constraint
is, the lower the slack (or surplus) for the constraint.)

(d) Can we obtain convex functions $f(x)$ for the functions mentioned in 
(\ref{eq:polyCvxFunc}) and (\ref{eq:fracPolyCvxFunc})?

A lot more testing is required for higher values of $N$ to show that this
method yields a polynomial time exact algorithm to find optimal solutions
for as many instances as possible (at least for some classes of MIS
instances).

\section{Conclusion}

We welcome readers to send us their comments and suggestions.

The latest version of this paper is usually posted at:
\newline
\tt{https://www.researchgate.net/publication/361555319}

Questions:

(a) In Step B of the 2-step procedure (Remark \ref{rem:twoStepProc}), for a
given $k$, if we are unable to find a feasible integer solution, can we
conclude that the optimal solution value to Problem (\ref{original_IP})
is less than $k$?  At this stage, with the algorithms reported in this
paper, we are still unable to do so.  Hopefully we can develop algorithms
with which we are able to make such a conclusion.

(b) Also in Step B, if the total cost (the objective function value) is the
same as the \it{desiredCost},
can we conclude that the solution obtained will be (or must be) an
\it{integer solution}, that is, a solution where for every vertex $i$,
$F[i]$ $\in$ $\{1, w\}$?  Again, with the algorithms in this paper,
we are unable to do so.

(\it{desiredCost} is defined in the models in Appendices 
\ref{sec:LPofStep1} and \ref{sec:MINOSprogram}.)

 \appendix

\section{Linear Program to determine the 9 parameters other than
$w$}\label{sec:LPofStep1}

We first assume a value for $w$.  The 9 other parameters whose values we
need (in order to solve MIS as a convex programming problem) are: 
$C$, $a_1$, $a_2$, $a_3$, $a_4$, $b_1$, $b_2$, $b_3$ and $b_4$. ~The LP
below is in AMPL format.. Any LP solver can be used (we used Gurobi).

\begin{verbatim}
param N := 25;
param k := 4;
param intvl := 100000;
param eps := 30;
# eps is a lower bound for the difference above desiredCost (how far
# above desiredCost a quantity should be).

param lowCurv := 500;
param curv_lower_bound := 0.00000001;

# We DON'T input "w" directly.. Instead, we input the values of lowCurv
# and intvl.. Usually we keep "intvl" as the same for all (or most)
# experiments.. From experiment to experiment, we only change the value
# of "lowCurv"..

# (Of course, from experiment to experiment, we may also change other
# input such as N, k, eps, and curv_lower_bound.)

param w := lowCurv/intvl;

param equalWt := w*(N-k)/N + k/N;
param Nkw12 := (N-k)*w/(N-12*k);
param Nkw8 := (N-k)*w/(N-8*k);
param Nkw6 := (N-k)*w/(N-6*k);
param Nkw5 := (N-k)*w/(N-5*k);
param Nkw4 := (N-k)*w/(N-4*k);
param Nkw3 := (N-k)*w/(N-3*k);
param Nkw2 := (N-k)*w/(N-2*k);

set fValues := 1..intvl;

var equalWtTotalCost;
var  C;
var a1;
var a2;
var a3;
var a4;
var b1;
var b2;
var b3;
var b4;
var Func {fValues};
var funcW;
var func1;
var Func2;
var Func3;
var Func4;
var Func5;
var Func6;
var Func8;
var Func12;
var Nkw12Diff;
var Nkw8Diff;
var Nkw6Diff;
var Nkw5Diff;
var Nkw4Diff;
var Nkw3Diff;
var Nkw2Diff;
var equalWeightFunc;
var equalWeightDiff;
var Nkw12Int;
var Nkw8Int;
var Nkw6Int;
var Nkw5Int;
var Nkw4Int;
var Nkw3Int;
var Nkw2Int;
var desiredCost;
var W2;
var W3;
var W4;
var W5;
var W10;
var W20;
var W100;
var W1000;
var V001999;
var V0199;
var V0298;
var V0595;
var V1585;
var V37;
var curvature {fValues};


# Objective function (actually NOT necessary, since we are just looking
# for a feasible solution to this LP):
maximize lkdjcdsfs: func1;

#  Just in case the objective function becomes unbounded, we set an upper
#  bound:
subject to func1Limit:
func1 <= 10000000000;

# We define the function f(x) here, where x = (f/intvl):
subject to function_def {f in fValues}:
a4*((f/intvl)^4) + a3*((f/intvl)^3) +
a2*((f/intvl)^2) + a1*f/intvl + C + b1*intvl/f + b2*((intvl/f)^2) 
+ b3*((intvl/f)^3) + b4*((intvl/f)^4) - Func[f] = 0;


# (Of course, instead of "pre-defining" the function at specific points,
# we can compute the function value as and when necessary.. For instance,
# if we need to encode the constraint F(x) - F(y) <= T, we would code it
# as below:)
#   a4*x^4 + a3*x^3 + a2*x^2 + a1*x + b4/x^4 + b3/x^3 + b2/x^2 + b1/x
# - a4*y^4 - a3*y^3 - a2*y^2 - a1*y - b4/y^4 - b3/y^3 - b2/y^2 - b1/y
#   <= T;


# Function value at (x = w):
subject to funcW_def:
a4*w*w*w*w + a3*w*w*w + a2*w*w + a1*w + C + b1/w + b2/(w*w) + b3/(w*w*w)
+ b4/(w*w*w*w) - funcW = 0;

# Function value at (x = 1.0):
subject to func1_def:
a4 + a3 + a2 + a1 + C + b1 + b2 + b3 + b4 - func1 = 0;

# desiredCost -- this is the cost of the solution that we desire -- which
# is, placing the k unit-sized items in k bins... and in the remaining
# (N-k) bins, place an item of size w.

subject to desiredFuncvalueDef:
k*func1 + (N-k)*funcW - desiredCost = 0;

# -----------------------------------

# The total weight of items available = k + (N-k)*w... If we divide this
# into N equally weighted items and place them one in each bin, the total
# cost of such a placement should be higher than desiredCost defined
# above.

subject to equalWeightFuncDef:
a4*equalWt*equalWt*equalWt*equalWt + a3*equalWt*equalWt*equalWt +
a2*equalWt*equalWt + a1*equalWt + C + b1/equalWt + b2/(equalWt*equalWt) +
b3/(equalWt*equalWt*equalWt) + b4/(equalWt*equalWt*equalWt*equalWt) -
equalWeightFunc = 0;

subject to equalWeightDef:
N*equalWeightFunc - equalWtTotalCost = 0;

subject to totalEqualWeightFuncDef:
equalWtTotalCost - desiredCost - equalWeightDiff = 0;

subject to break_6_pieces_6inverse_plus_w:
equalWeightDiff >= eps;

# -----------------------------------

# 12 here means, each unit-sized item is broken up into 12 equal parts,
# each of size (1/12).

subject to Nkw12_function:
a4*Nkw12*Nkw12*Nkw12*Nkw12 + a3*Nkw12*Nkw12*Nkw12 + a2*Nkw12*Nkw12 +
a1*Nkw12 + C + b1/Nkw12 + b2/(Nkw12*Nkw12) + b3/(Nkw12*Nkw12*Nkw12) +
b4/(Nkw12*Nkw12*Nkw12*Nkw12) - Nkw12Int = 0;

# Func12 = Func[1/12], when an item of size (1/12) is placed in
# its own bin.
subject to Func12def:
a4/20736 + a3/1728 + a2/144 + a1/12 + C + 
b1*12 + b2*144 + b3*1728 + b4*20736 - Func12 = 0;

subject to Nkw12Difference:
(N-12*k)*Nkw12Int + 12*k*Func12 - desiredCost - Nkw12Diff = 0;

# When the "if" condition is satisfied, set the constraint 
# Nkw12Diff >= eps ..... And when the "if" condition is NOT satisfied,
# DON'T set such a constraint:
subject to break_into_12_pieces:
Nkw12Diff >= if (N > 12*k and Nkw12 <= 1.0 and 0.083333 >= w) then eps;

# (N > 12*k) implies that Nkw12 > 0. 

# -----------------------------------

# 8 here means, each unit-sized item is broken up into 8 equal parts,
# each of size 0.125.

subject to Nkw8_function:
a4*Nkw8*Nkw8*Nkw8*Nkw8 + a3*Nkw8*Nkw8*Nkw8 + a2*Nkw8*Nkw8 +
a1*Nkw8 + C + b1/Nkw8 + b2/(Nkw8*Nkw8) + b3/(Nkw8*Nkw8*Nkw8) +
b4/(Nkw8*Nkw8*Nkw8*Nkw8) - Nkw8Int = 0;

# Func8 = = Func[1/8] = Func[0.125], when an item of size 0.125 is placed
# in its own bin.
subject to Func8def:
a4/4096 + a3/512 + a2/64 + a1/8 + C + 
b1*8 + b2*64 + b3*512 + b4*4096 - Func8 = 0;

subject to Nkw8Difference:
(N-8*k)*Nkw8Int + 8*k*Func8 - desiredCost - Nkw8Diff = 0;
  
# When the "if" condition is satisfied, set the constraint 
# Nkw8Diff >= eps ..... And when the "if" condition is NOT satisfied,
# DON'T set such a constraint:
subject to break_into_8_pieces:
 Nkw8Diff >= if (N > 8*k and Nkw8 <= 1.0 and 0.125 >= w) then eps;

# (N > 8*k) implies that Nkw8 > 0. 

# -----------------------------------

# 6 here means, each unit-sized item is broken up into 6 equal parts,
# each of size (1/6).

subject to Nkw6_function:
a4*Nkw6*Nkw6*Nkw6*Nkw6 + a3*Nkw6*Nkw6*Nkw6 + a2*Nkw6*Nkw6 +
a1*Nkw6 + C + b1/Nkw6 + b2/(Nkw6*Nkw6) + b3/(Nkw6*Nkw6*Nkw6) +
b4/(Nkw6*Nkw6*Nkw6*Nkw6) - Nkw6Int = 0;

subject to Func6def:
a4/1296 + a3/216 + a2/36 + a1/6 + C + 
b1*6 + b2*36 + b3*216 + b4*1296 - Func6 = 0;

subject to Nkw6Difference:
(N-6*k)*Nkw6Int + 6*k*Func6 - desiredCost - Nkw6Diff = 0;
  
# When the "if" condition is satisfied, set the constraint 
# Nkw6Diff >= eps ..... And when the "if" condition is NOT satisfied,
# DON'T set such a constraint:
subject to break_into_6_pieces:
 Nkw6Diff >= if (N > 6*k and Nkw6 <= 1.0 and 0.16667 >= w) then eps;

# (N > 6*k) implies that Nkw6 > 0. 

# -----------------------------------

# 5 here means, each unit-sized item is broken up into 5 equal parts,
# each of size 0.2.

subject to Nkw5_function:
a4*Nkw5*Nkw5*Nkw5*Nkw5 + a3*Nkw5*Nkw5*Nkw5 + a2*Nkw5*Nkw5 +
a1*Nkw5 + C + b1/Nkw5 + b2/(Nkw5*Nkw5) + b3/(Nkw5*Nkw5*Nkw5) +
b4/(Nkw5*Nkw5*Nkw5*Nkw5) - Nkw5Int = 0;

# Func5 = Func[1/5] = Func[0.2], when an item of size 0.2 is placed
# in its own bin.
subject to Func5def:
a4/625 + a3/125 + a2/25 + a1/5 + C + 
b1*5 + b2*25 + b3*125 + b4*625 - Func5 = 0;

subject to Nkw5Difference:
(N-5*k)*Nkw5Int + 5*k*Func5 - desiredCost - Nkw5Diff = 0;
  
subject to break_into_5_pieces:
 Nkw5Diff >= if (N > 5*k and Nkw5 <= 1.0 and 0.2 >= w) then eps;

# (N > 5*k) implies that Nkw5 > 0. 

# -----------------------------------

subject to Nkw4_function:
a4*Nkw4*Nkw4*Nkw4*Nkw4 + a3*Nkw4*Nkw4*Nkw4 + a2*Nkw4*Nkw4 +
a1*Nkw4 + C + b1/Nkw4 + b2/(Nkw4*Nkw4) + b3/(Nkw4*Nkw4*Nkw4) +
b4/(Nkw4*Nkw4*Nkw4*Nkw4) - Nkw4Int = 0;

# Func4 = Func[1/4] = Func[0.25], when an item of size 0.25 is placed
# in its own bin.

subject to Func4def:
a4/256 + a3/64 + a2/16 + a1/4 + C + 
b1*4 + b2*16 + b3*64 + b4*256 - Func4 = 0;

subject to Nkw4Difference:
(N-4*k)*Nkw4Int + 4*k*Func4 - desiredCost - Nkw4Diff = 0;
  
subject to break_into_4_pieces:
Nkw4Diff >= if (N > 4*k and Nkw4 <= 1.0 and 0.25 >= w) then eps;

# (N > 4*k) implies that Nkw4 > 0.

# -----------------------------------
  
subject to Nkw3_function:
a4*Nkw3*Nkw3*Nkw3*Nkw3 + a3*Nkw3*Nkw3*Nkw3 + a2*Nkw3*Nkw3 +
a1*Nkw3 + C + b1/Nkw3 + b2/(Nkw3*Nkw3) + b3/(Nkw3*Nkw3*Nkw3) +
b4/(Nkw3*Nkw3*Nkw3*Nkw3) - Nkw3Int = 0;

subject to Func3def:
a4/81 + a3/27 + a2/9 + a1/3 + C + 
b1*3 + b2*9 + b3*27 + b4*81 - Func3 = 0;

subject to Nkw3Difference:
(N-3*k)*Nkw3Int + 3*k*Func3 - desiredCost - Nkw3Diff = 0;
  
subject to break_into_3_pieces:
Nkw3Diff >= if (N > 3*k and Nkw3 <= 1.0 and 0.3333 >= w) then eps;

# -----------------------------------
  
subject to Nkw2_function:
a4*Nkw2*Nkw2*Nkw2*Nkw2 + a3*Nkw2*Nkw2*Nkw2 + a2*Nkw2*Nkw2 +
a1*Nkw2 + C + b1/Nkw2 + b2/(Nkw2*Nkw2) + b3/(Nkw2*Nkw2*Nkw2) +
b4/(Nkw2*Nkw2*Nkw2*Nkw2) - Nkw2Int = 0;

subject to Func2def:
a4/16 + a3/8 + a2/4 + a1/2 + C + 
b1*2 + b2*4 + b3*8 + b4*16 - Func2 = 0;

subject to Nkw2Difference:
(N-2*k)*Nkw2Int + 2*k*Func2 - desiredCost - Nkw2Diff = 0;

subject to break_into_2_pieces:
Nkw2Diff >= if (N > 2*k and Nkw2 <= 1.0 and 0.5 >= w) then eps;

# Suggestion: Instead of saying  Nkw2Diff >= eps   one can say 
#    Nkw2Diff >= Factor*desiredCost   where Factor is a value
#   more than or equal to 1.0.

# -----------------------------------

subject to break_into_001_999A:
Func[0.001*intvl] + Func[0.999*intvl] - func1 - V001999 = 0;

subject to break_into_001_999B:
V001999 >= if (0.001 >= w) then eps;

# Suggestion: Instead of saying   V001999 >= eps   one can say 
#   V001999 >= Factor*func1   where Factor is a value
#   more than or equal to 1.0.

subject to break_into_01_99A:
Func[0.01*intvl] + Func[0.99*intvl] - func1 - V0199 = 0;

subject to break_into_01_99B:
V0199 >= if (0.01 >= w) then eps;

subject to break_into_05_95A:
Func[0.05*intvl] + Func[0.95*intvl] - func1 - V0595 = 0;

subject to break_into_05_95B:
V0595 >= if (0.05 >= w) then eps;

subject to break_into_02_98A:
Func[0.02*intvl] + Func[0.98*intvl] - func1 - V0298 = 0;

subject to break_into_02_98B:
V0298 >= if (0.02 >= w) then eps;

subject to break_into_15_85A:
Func[0.15*intvl] + Func[0.85*intvl] - func1 - V1585 = 0;

subject to break_into_15_85B:
V1585 >= if (0.15 >= w) then eps;

subject to break_into_3_7A:
Func[0.3*intvl] + Func[0.7*intvl] - func1 - V37 = 0;

subject to break_into_3_7B:
V37 >= if (0.3 >= w) then eps;

subject to two_pieces_noEmptyBinsA:
2*Func[0.5*intvl] - func1 - W2 = 0;

subject to two_pieces_noEmptyBinsB:
W2 >= if (0.5 >= w) then eps;

subject to three_pieces_noEmptyBinsA:
3*Func[0.33333*intvl] - func1 - W3 = 0;

subject to three_pieces_noEmptyBinsB:
W3 >= if (0.33333 >= w) then eps;

subject to four_pieces_noEmptyBinsA:
4*Func[0.25*intvl] - func1 - W4 = 0;

subject to four_pieces_noEmptyBinsB:
W4 >= if (0.25 >= w) then eps;

subject to five_pieces_noEmptyBinsA:
5*Func[0.2*intvl] - func1 - W5 = 0;

subject to five_pieces_noEmptyBinsB:
W5 >= if (0.2 >= w) then eps;

subject to ten_pieces_noEmptyBinsA:
10*Func[0.1*intvl] - func1 - W10 = 0;

subject to ten_pieces_noEmptyBinsB:
W10 >= if (0.1 >= w) then eps;

subject to twenty_pieces_noEmptyBinsA:
20*Func[0.05*intvl] - func1 - W20 = 0;

subject to twenty_pieces_noEmptyBinsB:
W20 >= if (0.05 >= w) then eps;

subject to hundred_pieces_noEmptyBinsA:
100*Func[0.01*intvl] - func1 - W100 = 0;

subject to hundred_pieces_noEmptyBinsB:
W100 >= if (0.01 >= w) then eps;

subject to thousand_pieces_noEmptyBinsA:
1000*Func[0.001*intvl] - func1 - W1000 = 0;

subject to thousand_pieces_noEmptyBinsB:
W1000 >= if (0.001 >= w) then eps;

# ------- CONVEXITY CHECK (using second derivative) --------

subject to second_derivative {f in fValues}:
12*a4*((f/intvl)^2) + 6*a3*f/intvl + 2*a2 + 2*b1*((intvl/f)^3) +
6*b2*((intvl/f)^4) + 12*b3*((intvl/f)^5) +
20*b4*((intvl/f)^6) - curvature[f] = 0;

subject to curvature_condition {f in lowCurv..intvl}:
curvature[f] >= curv_lower_bound;

# ------- CONVEXITY CHECK ENDS --------

solve;

# The "display" command does NOT return precise output.. If you want
# precise output, use the "printf" command.

display N;
display k;
display C;
display a4;
display a3;
display a2;
display a1;
display b4;
display b3;
display b2;
display b1;
display w;
display desiredCost;
display equalWt;
display equalWeightFunc;
display equalWeightDiff;
display func1;
display funcW;
display Nkw12Diff;
display Nkw8Diff;
display Nkw6Diff;
display Nkw5Diff;
display Nkw4Diff;
display Nkw3Diff;
display Nkw2Diff;
display W2;
display W3;
display W4;
display W5;
display W10;
display W20;
display W100;
display W1000;
display V001999;
display V0199;
display V0298;
display V0595;
display V1585;
display V37;
display lowCurv;
display intvl;
display eps;
display curv_lower_bound;

\end{verbatim}

\section{Convex Program to solve M.I.S.}\label{sec:MINOSprogram}

We minimise a convex function over a convex space.  This was solved using
the MINOS solver at the NEOS server.
\it{All constraints are linear; hence the feasible region is a convex
space.}  (However, the function minimised is \bf{not} linear.)

\begin{verbatim}

option minos_options  ' \
    superbasics_limit=2000\
    scale_option=2\
    solution=yes\
 ';


param N;
param k;
param w >= 0, <= 1;
param C;
param a4;
param a3;
param a2;
param a1;
param b4;
param b3;
param b2;
param b1;

set Vertices := 1..N;
set Edges within (Vertices cross Vertices);

var X {Vertices} >= w, <= 1.0;
var Func {Vertices};
var desiredCost;
var totalCost;
var costDifference;
var Ftotal;
var func1;
var funcW;


minimize independent_set: totalCost;

subject to function_def {f in Vertices}:
a4*X[f]*X[f]*X[f]*X[f] + a3*X[f]*X[f]*X[f] + a2*X[f]*X[f] + a1*X[f] + C +
b1/X[f] + b2/(X[f]*X[f]) + b3/(X[f]*X[f]*X[f]) + b4*(X[f]*X[f]*X[f]*X[f])
- Func[f] = 0;

# CAUTION:  Variable "f" above refers to a graph vertex, NOT some value
# between zero and one !!

subject to totalCostDef:
totalCost - sum {f in Vertices} Func[f] = 0;

subject to Cost_difference_def:
costDifference = totalCost - desiredCost;

# Strictly, we need a constraint that says "costDifference = 0"
# but I have relaxed it a little:

subject to totalCost_Le_DesiredCost:
costDifference <= 1000;

subject to totalCost_Ge_DesiredCost:
costDifference >= -1000;

# Function value at x = w:
subject to funcW_def:
a4*w*w*w*w + a3*w*w*w + a2*w*w + a1*w + C + b1/w + b2/(w*w) + b3/(w*w*w) +
b4/(w*w*w*w) - funcW = 0;

# Function value at x = 1.0:
subject to func1_def:
a4 + a3 + a2 + a1 + C + b1 + b2 + b3 + b4 - func1 = 0;

subject to desiredFuncvalueDef:
k*func1 + (N-k)*funcW - desiredCost = 0;

subject to X_lowerbound {f in Vertices}:
X[f] >= w;

subject to X_upperbound {f in Vertices}:
X[f] <= 1.0;

# Note: Since we are adding an item of size "w" to the remaining N-k
# bins, the classical edge inequality "X[x] + X[y] <= 1" for edge (x, y)
# should be replaced by the following:

subject to independence {(x,y) in Edges}:
X[x] + X[y] <= 1+w;

subject to size_ind_set_constraintA:
sum {f in Vertices} X[f] = k + (N - k)*w;

data;

param N := 25;
param k := 4;

param C =  100000000.000082492828369;

param a4 = 0;

param a3 = 0;

param a2 =  0.000000005;

param a1 =  -0.000082507074949;

param b4 = 0;

param b3 = 0;

param b2 = 0;

param b1 = 0;

param w =  0.015;

# For each edge (i, j), the sum of the lengths of the items in the 2 bins
# i and j should be at most (1+w).. That is, Func[i] + Func[j] should be
# at most (1+w).

set Edges :=
(1, 3)
(1, 4)
(1, 5)
(1, 6)
(1, 7)
(1, 12)
(1, 14)
(1, 15)
(1, 16)
(1, 17)
(1, 18)
(1, 21)
(1, 22)
(1, 23)
(1, 24)
(2, 4)
(2, 5)
(2, 6)
(2, 7)
(2, 8)
(2, 9)
(2, 10)
(2, 12)
(2, 13)
(2, 15)
(2, 16)
(2, 18)
(2, 20)
(2, 22)
(2, 23)
(2, 24)
(2, 25)
(3, 6)
(3, 8)
(3, 9)
(3, 10)
(3, 11)
(3, 12)
(3, 14)
(3, 16)
(3, 17)
(3, 18)
(3, 19)
(3, 20)
(3, 21)
(3, 24)
(4, 6)
(4, 10)
(4, 11)
(4, 12)
(4, 13)
(4, 14)
(4, 15)
(4, 16)
(4, 17)
(4, 18)
(4, 19)
(4, 20)
(4, 21)
(4, 23)
(5, 6)
(5, 7)
(5, 9)
(5, 10)
(5, 11)
(5, 13)
(5, 14)
(5, 16)
(5, 17)
(5, 19)
(5, 20)
(5, 23)
(5, 24)
(6, 7)
(6, 8)
(6, 9)
(6, 12)
(6, 13)
(6, 16)
(6, 17)
(6, 18)
(6, 19)
(6, 20)
(6, 22)
(6, 23)
(6, 24)
(6, 25)
(7, 9)
(7, 11)
(7, 12)
(7, 13)
(7, 14)
(7, 17)
(7, 18)
(7, 20)
(7, 23)
(7, 25)
(8, 10)
(8, 11)
(8, 12)
(8, 15)
(8, 16)
(8, 17)
(8, 18)
(8, 20)
(8, 21)
(8, 22)
(8, 23)
(8, 24)
(9, 11)
(9, 12)
(9, 13)
(9, 15)
(9, 19)
(9, 20)
(9, 21)
(9, 22)
(9, 24)
(10, 11)
(10, 12)
(10, 14)
(10, 16)
(10, 17)
(10, 18)
(10, 21)
(10, 22)
(10, 23)
(10, 24)
(11, 14)
(11, 15)
(11, 17)
(11, 18)
(11, 19)
(11, 20)
(11, 21)
(11, 24)
(12, 13)
(12, 14)
(12, 16)
(12, 17)
(12, 19)
(12, 20)
(12, 21)
(12, 22)
(12, 23)
(12, 24)
(12, 25)
(13, 14)
(13, 15)
(13, 16)
(13, 17)
(13, 18)
(13, 19)
(13, 20)
(13, 21)
(13, 23)
(13, 24)
(13, 25)
(14, 17)
(14, 18)
(14, 19)
(14, 21)
(14, 22)
(15, 17)
(15, 19)
(15, 20)
(15, 24)
(15, 25)
(16, 17)
(16, 18)
(16, 20)
(16, 21)
(16, 23)
(16, 24)
(16, 25)
(17, 20)
(17, 21)
(17, 22)
(17, 23)
(17, 24)
(17, 25)
(18, 20)
(18, 21)
(18, 22)
(18, 24)
(18, 25)
(19, 20)
(19, 22)
(19, 24)
(19, 25)
(20, 21)
(20, 22)
(20, 24)
(20, 25)
(21, 22)
(21, 23)
(21, 25)
(22, 23)
(22, 25) ;

solve;

# The "display" command does NOT return precise output.. If you want
# precise output, use the "printf" command.

display X;
display Func;

display N;
display k;
display C;
display a4;
display a3;
display a2;
display a1;
display b4;
display b3;
display b2;
display b1;
display w;
printf "desiredCost= %d\n\n", desiredCost;
printf "totalCost= %d\n\n", totalCost;
printf "func1 = %d\n\n", func1;
printf "funcW = %d\n\n", funcW;

\end{verbatim}

\section{Modified LP model for Section \ref{sec:tighterConstraints}}

The original model is in Appendix A. ~We \bf{only} present the (modified
form of) the constraints in the original model that were modified.

\begin{verbatim}
subject to break_into_001_999A:
Func[floor((0.001 + w/2)*intvl)] + Func[floor((0.999 + w/2)*intvl)] -
func1 - funcW - V001999 = 0;

# You can use floor or ceiling, it doesn't matter, since we use intvl =
# 100,000, a fairly large number.

subject to break_into_001_999B:
V001999 >= if (0.001 >= w) then eps;

subject to break_into_01_99A:
Func[floor((0.01 + w/2)*intvl)] + Func[floor((0.99 + w/2)*intvl)] - func1
- funcW - V0199 = 0;

subject to break_into_01_99B:
V0199 >= if (0.01 >= w) then eps;

subject to break_into_05_95A:
Func[floor((0.05 + w/2)*intvl)] + Func[floor((0.95 + w/2)*intvl)] - func1
- funcW - V0595 = 0;

subject to break_into_05_95B:
V0595 >= if (0.05 >= w) then eps;

subject to break_into_02_98A:
Func[floor((0.02 + w/2)*intvl)] + Func[floor((0.98 + w/2)*intvl)] - func1
- funcW - V0298 = 0;

subject to break_into_02_98B:
V0298 >= if (0.02 >= w) then eps;

subject to break_into_15_85A:
Func[floor((0.15 + w/2)*intvl)] + Func[floor((0.85 + w/2)*intvl)] - func1
- funcW - V1585 = 0;

subject to break_into_15_85B:
V1585 >= if (0.15 >= w) then eps;

subject to break_into_3_7A:
Func[floor((0.3 + w/2)*intvl)] + Func[floor((0.7 + w/2)*intvl)] - func1 -
funcW - V37 = 0;

subject to break_into_3_7B:
V37 >= if (0.3 >= w) then eps;

subject to two_pieces_noEmptyBinsA:
2*Func[floor((0.5 + w/2)*intvl)] - func1 - funcW - W2 = 0;

subject to two_pieces_noEmptyBinsB:
W2 >= if (0.5 >= w) then eps;

subject to three_pieces_noEmptyBinsA:
3*Func[floor((1/3 + 2*w/3)*intvl)] - func1 - 2*funcW - W3 = 0;

subject to three_pieces_noEmptyBinsB:
W3 >= if (0.33333 >= w) then eps;

subject to four_pieces_noEmptyBinsA:
4*Func[floor((1/4 + 3*w/4)*intvl)] - func1 - 3*funcW - W4 = 0;

subject to four_pieces_noEmptyBinsB:
W4 >= if (0.25 >= w) then eps;

subject to five_pieces_noEmptyBinsA:
5*Func[floor((1/5 + 4*w/5)*intvl)] - func1 - 4*funcW - W5 = 0;

subject to five_pieces_noEmptyBinsB:
W5 >= if (0.2 >= w) then eps;

subject to ten_pieces_noEmptyBinsA:
10*Func[floor((1/10 + 9*w/10)*intvl)] - func1 - 9*funcW - W10 = 0;

subject to ten_pieces_noEmptyBinsB:
W10 >= if (0.1 >= w) then eps;

subject to twenty_pieces_noEmptyBinsA:
20*Func[floor((1/20 + 19*w/20)*intvl)] - func1 - 19*funcW - W20 = 0;

subject to twenty_pieces_noEmptyBinsB:
W20 >= if (0.05 >= w) then eps;

subject to hundred_pieces_noEmptyBinsA:
100*Func[floor((1/100 + 99*w/100)*intvl)] - func1 - 99*funcW - W100 = 0;

subject to hundred_pieces_noEmptyBinsB:
W100 >= if (0.01 >= w) then eps;

# subject to thousand_pieces_noEmptyBinsA:
# 1000*Func[floor((1/1000 + 999*w/1000)*intvl)] - func1 - 999*funcW -
# W1000 = 0;

# subject to thousand_pieces_noEmptyBinsB:
# W1000 >= if (0.001 >= w) then eps;

\end{verbatim}


\begin{thebibliography}{FP93}

\bibitem[FP93]{fang1993Linear}
Shu-Cherng Fang and Sarat Puthenpura.
\newblock {\em Linear optimization and extensions: theory and algorithms}.
\newblock Prentice-Hall, Inc., 1993.

\bibitem[GJ79]{gj}
M.R. Garey and D.S. Johnson.
\newblock {\em Computers and Intractability: A Guide to the Theory of
   NP-Completeness}.
\newblock Freeman (New York), 1979.

\bibitem[GH62]{hadley}
George Hadley.
\newblock {\em Linear Programming}.
\newblock Addison Wesley, 1962.

\end{thebibliography}
\end{document}